\documentclass[aps,prl,floatfix,showpacs,twocolumn]{revtex4}
\usepackage{amsmath,amssymb}
\usepackage{epsfig}

\newcommand{\bm}{\boldsymbol}

\begin{document}

%\twocolumn[
\hsize\textwidth\columnwidth\hsize\csname@twocolumnfalse\endcsname

\title{Quantum Anomalous Hall Effect in Single-layer and
  Bilayer Graphene}

\author{Wang-Kong Tse$^1$}
\author{Zhenhua Qiao$^1$}
\author{Yugui Yao$^{1,2}$} 
\author{A. H. MacDonald$^1$} 
\author{Qian Niu$^{1,3*}$} 
\affiliation{$^1$Department of Physics, University of Texas, Austin, Texas
  78712, USA}
\affiliation{$^2$Institute of Physics, Chinese Academy of Sciences,
  Beijing 100190, China}
\affiliation{$^3$International Center for Quantum Materials, Peking University, Beijing 100871, China}

\begin{abstract}
The quantum anomalous Hall effect can occur in 
single and few layer graphene systems that have both 
exchange fields and spin-orbit coupling. 
In this paper, we present a study of the quantum anomalous Hall effect in 
single-layer and gated bilayer graphene systems with Rashba spin-orbit coupling.  We compute Berry
curvatures at each valley point and find that for single-layer graphene the Hall conductivity is quantized 
at $\sigma_{xy} = 2e^2/h$, with each valley contributing
a unit conductance and a corresponding chiral edge state. 
In bilayer graphene, we find that the
quantized anomalous Hall conductivity is twice that of the 
single-layer case when the gate voltage $U$ is smaller than the
exchange field $M$, and zero otherwise. Although the Chern number 
vanishes when $U > M$, the system still exhibits a quantized valley Hall
effect, with the edge states in opposite valleys propagating in 
opposite directions.  The possibility of tuning between different topological states 
with an external gate voltage suggests possible  
graphene-based spintronics applications. 
\end{abstract}

%63.22.-m,63.20.D-,63.20.kd,71.10.-w}

\maketitle
%\newpage

\section{I. Introduction}

Graphene is a two-dimensional material with a single-layer 
honeycomb lattice of carbon atoms.  Its isolation in the past decade
has generated 
substantial theoretical and
experimental research activity \cite{Graphene_RMP}.   Experimental
fabrication methods continue to progress, motivating 
a recent focus on the possibility of utilizing graphene as a  
material for nanoelectronics.  At the same time, 
spintronics has also progressed in recent years. 
The spin degrees of freedom can be manipulated to encode information, 
allowing fast information processing and immense storage 
capacity \cite{Spintronics_RMP}.
This article is motivated by recent research progress which has advanced the 
prospects for spintronics in graphene.
 
A key element of spintronics is the 
presence of spin-orbit coupling which allows the spin degrees of
freedom to be controlled by electrical means.  It was pointed out some time ago that 
the quantum spin Hall effect \cite{Kane_Mele} can occur in a single-layer graphene 
sheets because of its intrinsic
spin-orbit (SO) coupling.  Instrinsic SO coupling 
%does not mix the two spin species, but 
induces momentum-space Berry curvatures (which act 
like momentum-space
magnetic fields)  that have opposite sign for
opposite spin.
However, the intrinsic coupling strength was later found to be
weak ($\sim 10^{-2}\,-\,10^{-3}\,\mathrm{meV}$ \cite{ISO_MacDonald,ISO_Fang,ISO_Fabian})
enough to make applications of the effect appear impractical.
Luckily another type of spin-orbit interaction known as Rashba SO coupling\cite{FirstRashbaSO}  appears when inversion
symmetry in the graphene plane is broken.
This SO coupling mixes different spins, so the spin-component perpendicular to the graphene plane 
is no longer conserved.  When acting alone, Rashba coupling causes the resulting spin 
eigenstates to be chiral.  One appealing feature of 
Rashba SO coupling is its tunability by 
an applied gate field $E_{\mathrm{G}}$.   Unfortunately the field-effect Rashba coupling strength 
is also weak ( $\sim 10\,-\,100\,\mu\mathrm{eV}$ per
$V/\mathrm{nm}$ \cite{ISO_MacDonald,ISO_Fabian}) at practical field 
strengths.  Recent experiments
\cite{ARPES_PRL_Rader,ARPES_PRL_Shikin} and ab initio calculation
\cite{TBpaper} have however suggested that surface deposition of 
impurity adatoms can dramatically enhance the Rashba SO coupling 
strength in graphene to $\sim 1\,-\,10\,\mathrm{meV}$,
raising the hope that spin transport effects induced
by Rashba SO coupling might be realizable as experiment
progresses.

The quantum anomalous Hall effect (QAHE) is characterized by a quantized charge Hall conductance in an
insulating state. Unlike the quantum Hall effect, which arises from 
Landau level quantization in a strong magnetic field, QAHE is induced
by internal magnetization and SO coupling. 
Although there have been a number of theoretical studies of this
unusual effect \cite{QAHEpapers1,QAHEpapers2,QAHEpapers3,QAHEpapers4,QAHEpapers5,QAHEpapers6}, no experimental observation has been reported so far.
In a recent article \cite{TBpaper}, we predicted on the basis of
tight-binding lattice models and {\em ab initio} calculations that QAHE can occur in single-layer graphene in the presence of both
an exchange field and Rashba SO coupling. 
In this paper, we complement our previous numerical work with a 
continuum model study which yields more analytical progress and provides clearer insight into our findings.
We also present a more detailed and systematic investigation of the topological phases 
of both single-layer and gated bilayer graphene with strong Rashba SO interactions.
In single-layer graphene the Hamiltonian is
analytically diagonalizable and we obtain an analytic expression for the 
Berry curvature and use it to evaluate the Chern number. For 
bilayer graphene, the possibility of a gate field applied across the
bilayer introduces a tunable parameter which we show can 
induce a topological phase transition. We find
that when the bilayer potential difference $U$ is smaller than the
exchange field $M$, the system behaves as a quantum anomalous Hall
insulator, whereas for $U > M$, the system behaves as a quantum valley
Hall insulator with zero Chern number. For each case, we also study the edge state
properties of the corresponding finite system using a numerical tight-binding approach. 

The paper is organized as follows. We first discuss the bulk
topological properties for the case of
single-layer graphene in section II. In section III, we turn our
attention to the case of gated bilayer graphene. We then discuss the
edge state properties of both the single-layer and the gated bilayer
cases in section IV.  Our conclusion are presented in section V. An appendix follows
that develops an envelope function analysis of the edge states in the
single-layer system.

\section{II. Single-layer Graphene}

The Brillouin zone of graphene is hexagonal with two inequivalent K
and K' points located at the zone corners. The band structure has a linear band crossings
at both K and K' .  At wavevectors near either of these valley points,
the envelope-function wavefunctions satisfy a massless Dirac equation. We represent the graphene
envelope-function Hamiltonian in the basis
$\{\mathrm{A}\uparrow,\mathrm{B}\uparrow,\mathrm{A}\downarrow,\mathrm{B}\downarrow\}$
for both valleys K and K'.
%for valley K and $\{\mathrm{B}\uparrow,\mathrm{A}\uparrow,\mathrm{B}\downarrow,\mathrm{A}\downarrow\}$
%for valley K'. 

Rashba SO coupling in graphene was first discussed by Kane and Mele \cite{Kane_Mele} and subsequently
by Rashba \cite{Rashba}, and also 
%by a number of other authors 
in a number of recent papers \cite{Sandler,Smith,Wrinkler}. 
The Hamiltonian for valley K including Rashba SO coupling and exchange
field is 
\begin{equation}
H =v\bm{\sigma}\cdot\bm{k}\bm{1}_s+\alpha\left(\bm{\sigma}\times\bm{s}\right)_z+M
\bm{1}_{\sigma}s_z,
%H =\zeta v\bm{\sigma}\cdot\bm{k}\bm{1}_s+\zeta \alpha\left(\bm{\sigma}\times\bm{s}\right)_z+M
%\bm{1}_{\sigma}s_z,
\label{eq1}
\end{equation}
where $\bm{\sigma}$ and $\bm{s}$ are Pauli matrices that correspond
respectively to the pseudospin (i.e., A-B
sublattice) and spin degrees of freedom, $\bm{1}_{\sigma,s}$ denotes
the identity matrix in the $\sigma$ and $s$ space, $\alpha$ is the Rashba SO coupling
strength, and $M$ is the exchange magnetization. The Hamiltonian for
valley K' is obtained by the replacement $\bm{\sigma}\to -\bm{\sigma}^*$.
%and $\zeta = \pm 1$ corresponds to valley K and K' respectively. 
We neglect intrinsic
SO coupling since we are interested in the
case when the Rashba SO coupling parameter $\alpha$ is much stronger than the intrinsic 
coupling parameter  $\Delta_{\mathrm{intrinsic}}$.  We note, however, that the presence of a small but finite
intrinsic SO coupling is not expected to qualitatively modify our
results as long as $\Delta_{\mathrm{intrinsic}} \ll M$.
%as the
%zero-energy crossings between spin up and spin down states still remain intact with only $\Delta_{\mathrm{intrinsic}}$ and $M$. 
%
\begin{figure}[t]
  \includegraphics[width=8.5cm,angle=0]{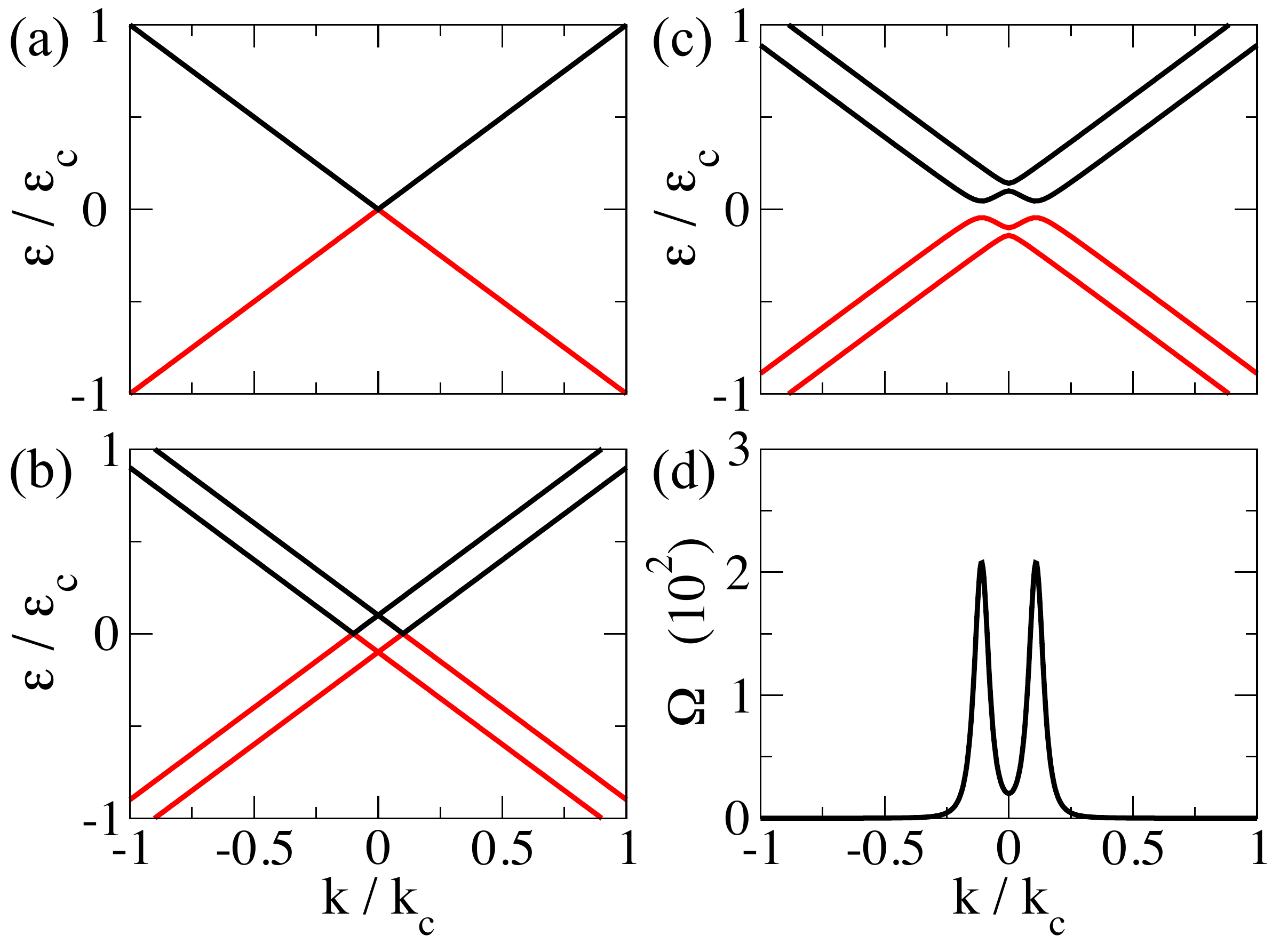}
\caption{(Color online) (a)-(c) Bulk states band structure. (a). $M
  =0$, $\alpha = 0$, (b). $M/\varepsilon_{\mathrm{c}} = 0.1$, $\alpha
  = 0$, (c). $M/\varepsilon_{\mathrm{c}} = 0.1$, $\alpha/\varepsilon_{\mathrm{c}} =
  0.05$. $k_{\mathrm{c}} = 2\pi/a$ ($a$ is the
graphene lattice constant) and $\varepsilon_{\mathrm{c}}$ are momentum
and energy cut-off of the Dirac model, beyond which the energy dispersion deviates
from linearity due to trigonal warping. (d). Berry curvature $\Omega = 2(\Omega_{+-}+\Omega_{--})$
  (the factor of two arises from the two valleys) for $\alpha/\varepsilon_{\mathrm{c}} 
= 0.05$, $M/\varepsilon_{\mathrm{c}}  = 0.1$. $\Omega$ peaks at the $k$ value where the upper valance band 
has its maximum and degeneracy between opposite spin states occurs 
when SO coupling is absent [see panel (b)].} \label{BandBerry}
\end{figure}

Upon diagonalization of the Hamiltonian, we obtain the energy
dispersion for both valleys
\begin{equation}
\varepsilon_{ks\mu} =
\mu\sqrt{M^2+\epsilon_k^2+2\alpha^2+2s\sqrt{\alpha^4+\epsilon_k^2\left(M^2+\alpha^2\right)}},
\label{eq2}
\end{equation}
where $\epsilon_k = v k$, $\mu = \pm$ stands for the conduction $(+)$ and valence $(-)$
bands. 
%, and $s = \pm$ is the spin chirality label. 
Because of spin-mixing due to Rashba SO coupling, 
spin is no longer a good quantum number, and the resulting
angular momentum eigenstates are denoted by the spin chirality $s = \pm$. 
The band structure therefore consists of two spin chiral conduction
bands and two spin chiral valence bands. The corresponding eigenstates are 
%
%\begin{widetext}
\begin{eqnarray}
&&u_{s\mu}\left(k\right) = \label{eq3} \\
&&N_{s\mu}\left[\begin{array}{cccc}
\zeta i e^{-i2\phi_k}P_{s\mu}, &
%\left[-M\epsilon_k^2+\left(\alpha^2-s\sqrt{\alpha^4+\epsilon_k^2\left(M^2+\alpha^2\right)}\right)\left(M+\varepsilon_{ks\mu}\right)\right] \\
i e^{-i\phi_k}Q_{s\mu}, &
%\epsilon_k\left[\epsilon_k^2-\left(M+\varepsilon_{ks\mu}\right)^2\right]/2
%\\
\zeta e^{-i\phi_k} R_{s\mu}, &
%\alpha\epsilon_k\left(M+\varepsilon_{ks\mu}\right) \\
\alpha\epsilon_k^2
\end{array}\right]^{\mathrm{T}}, \nonumber
\end{eqnarray}
%\end{widetext}
%
where $\phi_k = \tan^{-1}(k_y/k_x)$, $N_{s\mu}$ is the normalization constant
%
%\begin{widetext}
%\begin{eqnarray}
\begin{equation}
N_{s\mu}\left(k\right) =\left\{P_{s\mu}^2+Q_{s\mu}^2+R_{s\mu}^2+\left(\alpha\epsilon_k^2\right)^2\right\}^{-1/2},
% \left\{\left[-M\epsilon_k^2+\left(\alpha^2-s\sqrt{\alpha^4+\epsilon_k^2\left(M^2+\alpha^2\right)}\right)\left(M+\varepsilon_{ks\mu}\right)\right]^2\right.
%   \nonumber \\
%&&\left.+\epsilon_k^2\left[\epsilon_k^2-\left(M+\varepsilon_{ks\mu}\right)^2\right]^2/4+\alpha^2\epsilon_k^2\left(M+\varepsilon_{ks\mu}\right)^2+\alpha^2\epsilon_k^%4\right\}^{-1/2}. 
\label{eq4}
\end{equation}
%\end{eqnarray}
%\end{widetext}
%
and $P, Q, R$ are functions defined as follows:
\begin{eqnarray}
&&P_{s\mu}\left(k\right) = \nonumber \\
&&-M\epsilon_k^2+\left(\alpha^2-s\sqrt{\alpha^4+\epsilon_k^2\left(M^2+\alpha^2\right)}\right)\left(M+\varepsilon_{ks\mu}\right),
\nonumber \\
&&Q_{s\mu}\left(k\right) =
\epsilon_k\left[\epsilon_k^2-\left(M+\varepsilon_{ks\mu}\right)^2\right]/2,
\nonumber \\
&&R_{s\mu}\left(k\right) = \alpha\epsilon_k\left(M+\varepsilon_{ks\mu}\right).
\label{eq5}
\end{eqnarray} 

Fig.~\ref{BandBerry} illustrates the evolution of the electronic structure as the exchange field $M$ and Rashba SO coupling $\alpha$ are introduced
to the system. As shown in Fig.~\ref{BandBerry}(b), the exchange field splits the original spin-degenerate
Dirac cone into two oppositely spin-polarized copies, and this in turn produces spin degeneracy circles in momentum space 
at energy $\varepsilon = 0$. 
Introducing the SO coupling $\alpha$ causes a gap to open up between the conduction and valence bands around this circle 
along which SO coupling 
mixes up and down spins and produces an avoided band crossing. 
The momentum magnitude $k =  k_{\mathrm{\Delta}}$ at which the avoided crossing occurs and the gap $\Delta$ are given by 
\begin{equation}
k_{\mathrm{\Delta}} = \frac{M \sqrt{M^2 + 2 \alpha^2}}{v \sqrt{M^2 +
    \alpha^2}},\,\,\,\,\,\Delta = \frac{2 \alpha M}{\sqrt{M^2 + \alpha^2}}.
\label{kgap}
\end{equation}
%
%and 
%
%\begin{equation}
%\Delta = \frac{2 \alpha M}{\sqrt{M^2 + \alpha^2}}.
%\label{Gap}
%\end{equation}

In the insulating regime when the Fermi level lies within the 
bulk gap, the Hall conductivity $\sigma_{xy} = \mathcal{C}e^2/h$ where $\mathcal{C}$ 
is the Chern number $\mathcal{C}$ which can be evaluated using the 
Thouless-Kohmoto-Nightingale-den Nijs (TKNN's) formula \cite{BerryPhaseRMP}:
\begin{equation}
\mathcal{C} = \frac{1}{2\pi}\sum_n\int{\mathrm{d}^2k}\left({\bm{\Omega}_n}\right)_z,
\label{ChernNumber}
\end{equation}
where $n$ labels the bands below the Fermi level, and $\bm{\Omega}_n$ is the Berry curvature
\begin{equation}
\bm{\Omega}_n = i\langle\frac{\partial
  u_n}{\partial\bm{k}}\vert\times\vert \frac{\partial
  u_n}{\partial\bm{k}}\rangle,
\label{BerryC}
\end{equation}
with $u_n$ denoting the Bloch state for band $n$. Before calculating the Chern number, we briefly comment on and make connections
with two other formulas in the literature that are also used to calculate the Hall
conductivity in the insulating regime. 

For two-band Hamiltonians that can be written in the form $H =
\bm{\sigma}\cdot\bm{d}$, %with SU(2) symmetry, 
the TKNN formula can be
written in the form 
\begin{equation}
\mathcal{C} = \frac{1}{4\pi}\int\mathrm{d}^2k\left(\frac{\partial\bm{\hat{d}}}{\partial
  k_x}\times\frac{\partial \bm{\hat{d}}}{\partial k_y}\right)\cdot\bm{\hat{d}},
\label{winding1}
\end{equation}
where $\bm{\hat{d}} = \bm{d}/\vert \bm{d}\vert$ is the unit vector which specifies the 
direction of $\bm{d}$. The right hand side of Eq.~(\ref{winding1}) can be identified as a Pontryagin index which
is equal to the number of times the unit sphere of spinor directions is covered upon integrating over the 
Brillouin zone.  For the present case, however, 
the graphene Hamiltonian contains both spin and pseudospin degrees of
freedom, and Eq.~(\ref{winding1}) is not applicable. In this case, $\mathcal{C}$
is given by the following more general form of the Pontryagin index \cite{Volovik}:
\begin{equation}
\mathcal{C} =
\frac{1}{24\pi^2}\epsilon_{\mu\nu\lambda}\mathrm{tr}\int\mathrm{d}\omega\mathrm{d}^2k\;G\frac{\partial
G^{-1}}{\partial k_{\mu}}G\frac{\partial G^{-1}}{\partial
k_{\nu}}G\frac{\partial G^{-1}}{\partial k_{\lambda}},
\label{winding2}
\end{equation}
where $k_{\mu} = (\omega,k_x,k_y)$, $\epsilon_{\mu\nu\lambda}$ is the
anti-symmetric tensor, and $G = (i\omega-H)^{-1}$ is the Green 
function. In the non-interacting limit we consider in this work,
Eq.~(\ref{winding2}) can be shown to be equivalent \cite{Yakovenko} 
to the TKNN formula Eq.~(\ref{ChernNumber}). 

We now evaluate the $z$ component of the Berry curvature 
$\Omega_{s\mu}$ from Eq.~(\ref{BerryC}) for the bands which are labeled by $s$ and $\mu$.  
For each valley, we find that the Berry curvature is 
analytically expressible in terms of an exact differential
\begin{equation}
{\Omega}_{s\mu} = \frac{1}{k}\frac{\partial}{\partial
  k}\left[N_{s\mu}^2\left(2P_{s\mu}^2+Q_{s\mu}^2+R_{s\mu}^2\right)\right],
\label{Berry2}
\end{equation}
and the Chern number per valley for each band is given by
\begin{equation}
\mathcal{C}_{s\mu} 
%&=&
%\frac{1}{2\pi}\int{\mathrm{d}^2k}{\Omega}_{s\mu}
%\nonumber \\
%&=&
= N_{s\mu}^2\left[2P_{s\mu}^2+Q_{s\mu}^2+R_{s\mu}^2\right]\big\vert_{k
  = 0}^{k = \infty},
\label{Chern2}
\end{equation}
The upper limits of the integrand in Eq.~(\ref{Chern2}) can be set to $\infty$ 
because the Berry curvature is large only close to valley centers. 

Computing Eq.~(\ref{Chern2}), we find that in this continuum model the Chern number
$\mathcal{C}_{s-}$ for the individual valence band with $s = \pm$ is not quantized but
instead depends numerically on the specific values of
$\alpha$ and $M$. We find, however, that the two contributions always sum to $1$ and
therefore each valley carries a unit Chern number. 
Taking into account both valleys, it follows that the quantized Hall conductivity is
\begin{equation}
\sigma_{xy} = 2\frac{e^2}{h}\;\mathrm{sgn}(M).
\label{Cond}
\end{equation} 

\section{III. Gated Bilayer Graphene}

\begin{figure}[!tb]%[h]
  \includegraphics%[width=9cm,angle=0]
[width=8.5cm,angle=0]
{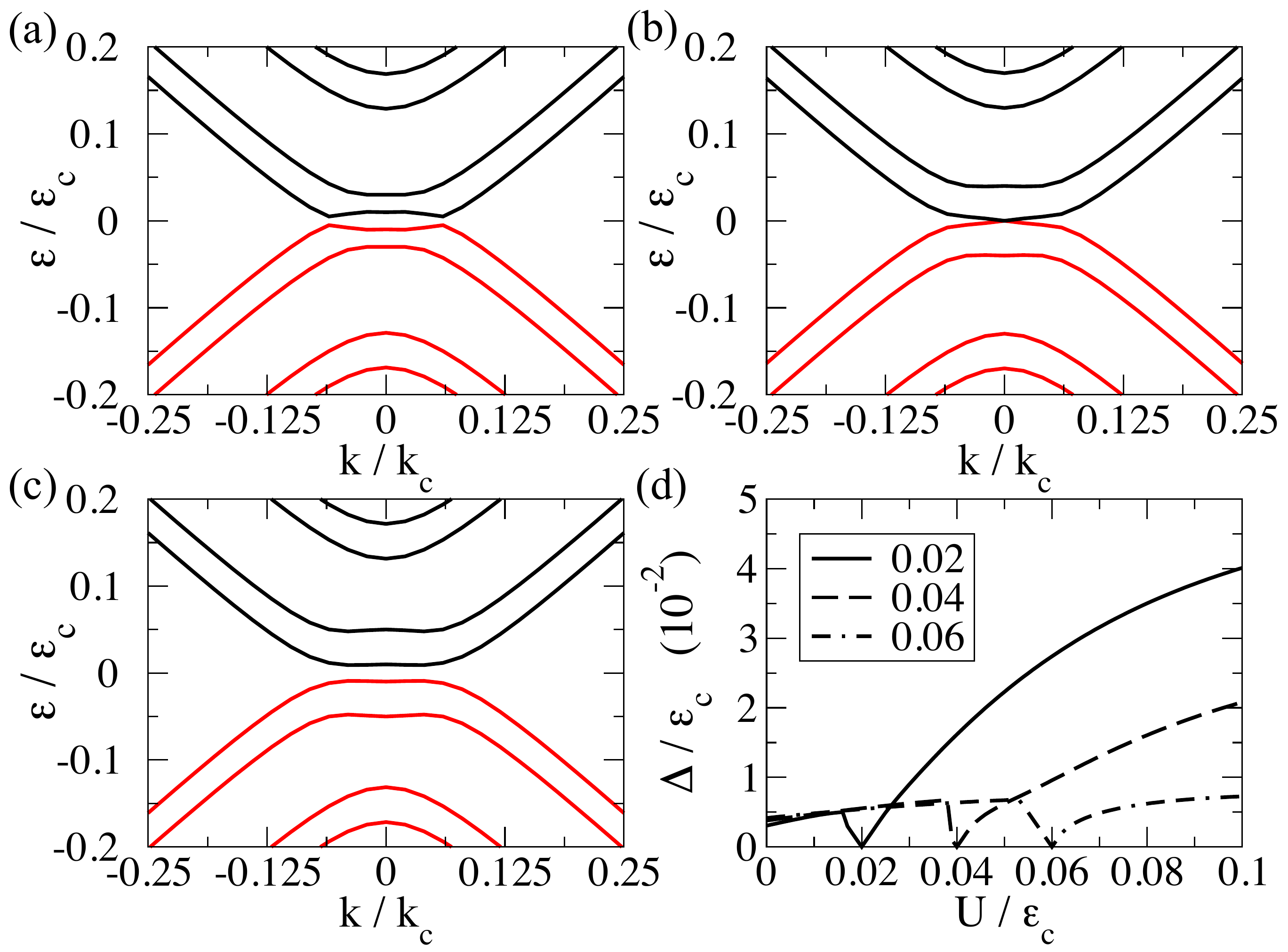} 
\caption{(Color online) Band structure at
  $\alpha/\varepsilon_{\mathrm{c}} 
= 5\times10^{-3}$ and $M/\varepsilon_{\mathrm{c}} = 0.02$ for (a)
$U/\varepsilon_{\mathrm{c}} = 0.01$, (b) $U/\varepsilon_{\mathrm{c}} =
0.02$, and (c) $U/\varepsilon_{\mathrm{c}} = 0.03$. (d) Band gap as a function of $U$ at
$\alpha/\varepsilon_{\mathrm{c}}  = 0.005$ and $M/\varepsilon_{\mathrm{c}} = 0.02, 0.04, 0.06$ as shown in legend.}
\label{fig0}
\end{figure}
We extend our discussion to the case of bilayer graphene. 
In the vicinity of valley K, we can write the bilayer graphene Hamiltonian in the presence of Rashba SO
coupling $\alpha$, exchange field $M$ and potential imbalance $U$ as ($\bm{\tau}$ denotes Pauli matrices for the layer degrees of freedom):
\begin{eqnarray}
H &=&
\left[v\bm{\sigma}\cdot\bm{k}1_s+M1_{\sigma}s_z+\left(\frac{\alpha_{\mathrm{T}}+\alpha_{\mathrm{B}}}{2}\right)\left(\bm{\sigma}\times\bm{s}\right)_z\right]1_{\tau}
\nonumber \\
&&+\left[\left(\frac{\alpha_{\mathrm{T}}-\alpha_{\mathrm{B}}}{2}\right)\left(\bm{\sigma}\times\bm{s}\right)_z+U1_{\sigma}1_s\right]\tau_z
\nonumber \\
&&+\frac{1}{2}t_{\perp}1_s\left(\sigma_x\tau_x+\sigma_y\tau_y\right),
\label{H}
\end{eqnarray}
where $t_{\perp} = 0.4\,\mathrm{eV}$ is the $\tilde{\mathrm{A}}\mathrm{B}$
interlayer hopping energy \cite{Mccann}. For the other valley K', the Hamiltonian is given by the
above with $\bm{\sigma}\to -\bm{\sigma}^*$. For generality, we have
written Eq.~(\ref{H}) allowing for different Rashba SO coupling strengths $\alpha_{\mathrm{T}}$
and $\alpha_{\mathrm{B}}$ for the top and bottom layers. We shall now set 
$\alpha_{\mathrm{T}} = \alpha_{\mathrm{B}} = \alpha$ for simplicity as the
specific values of $\alpha$ on the two layers do not alter the
topology of the bands in our discussions below. 
\begin{figure}[!tb]%[h]
  \includegraphics[width=6.5cm,angle=0]{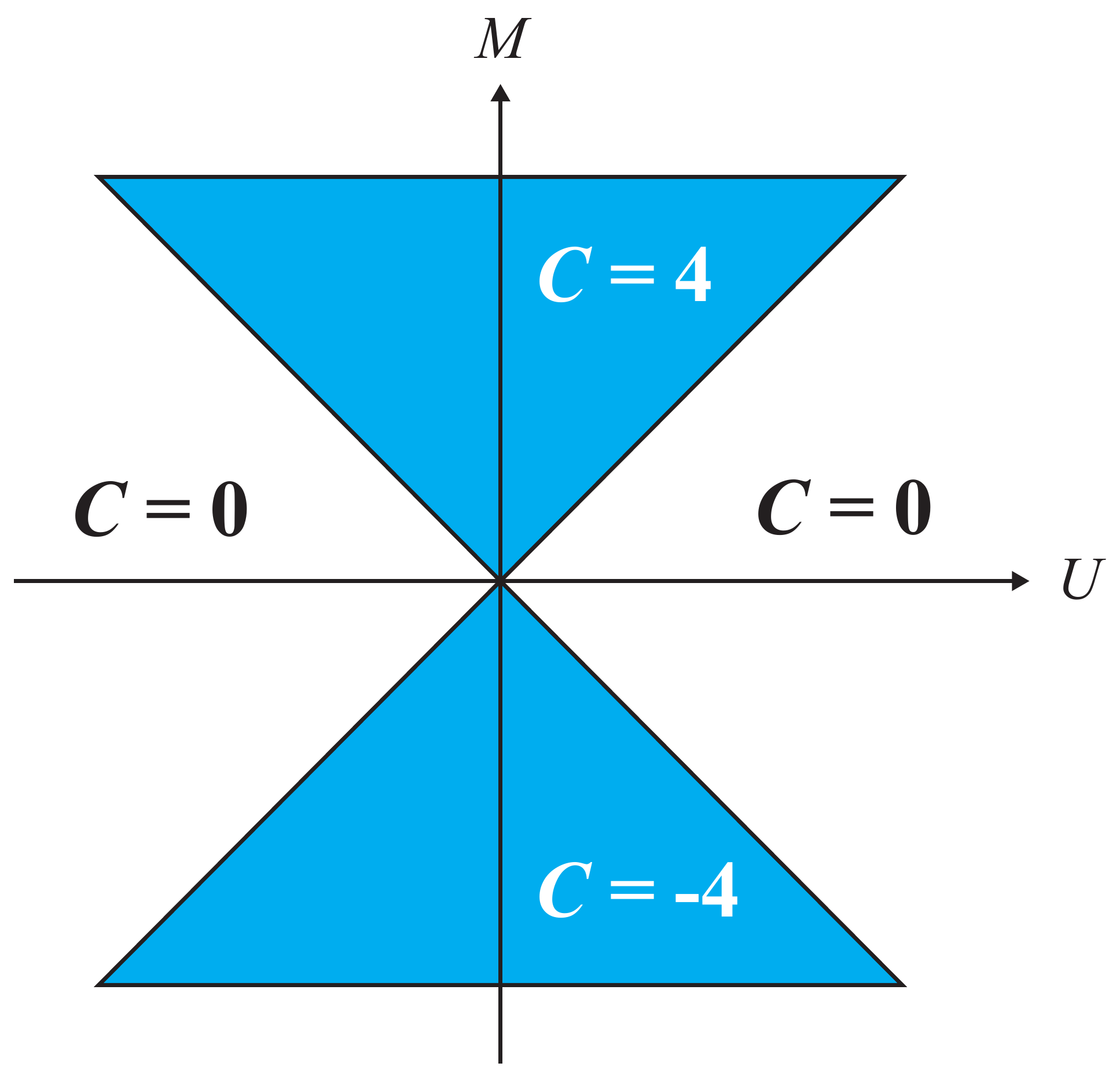} 
\caption{(Color online) Phase diagram of the Chern number
  $\mathcal{C}$ as a
  function of $U$ and $M$. The valley Chern number occupies 
  complementary regions of the phase space with $\mathcal{C}_v = 0$
  for $U < M$ and $4\mathrm{sgn}(U)$ for $U > M$.}
\label{fig1}
\end{figure}

The Hamiltonian Eq.~(\ref{H}) is not diagonalizable analytically \cite{Remark}. We
therefore obtain the eigenenergies and eigenvectors numerically and
use these to compute the Berry curvature. Fig.~\ref{fig0} shows the
band structure evolution for increasing values of $U$: $U < M$ (panel
a), $U = M$ (panel b), and $U > M$ (panel c). For $U < M$, an
inverted-gap profile appears that is similar to the single-layer graphene case
(Fig.~\ref{BandBerry}c). At $U = M$, the gap closes exactly at $k =
0$, and reopens when $U > M$. Fig.~\ref{fig0} shows the behavior of the
gap $\Delta$ as a function of the potential difference $U$ for various values of
$M$. We find that $\Delta$ initially increases with $U$ and then
decreases toward zero when $U$ approaches the value of $M$, after
which $\Delta$ increases again with $U$. 

The Berry curvature Eq.~(\ref{BerryC}) can be expressed in a form that is more 
convenient for numerical computation. For the $n^{\mathrm{th}}$ band,
the Berry curvature per valley can be expressed as 
\begin{equation}
\Omega_{xy}^n = -2\sum_{n' \neq n} \frac{\mathrm{Im}\left\{\langle n \vert v_x \vert n' \rangle
\langle n' \vert v_y \vert n \rangle\right\}}{\left(\varepsilon_n-\varepsilon_{n'}\right)^2}.
%
%i\sum_{n' \neq n} \frac{\langle n \vert v_x \vert n' \rangle
%\langle n' \vert v_y \vert n \rangle-\langle n \vert v_y \vert n' \rangle
%\langle n' \vert v_x \vert n
%\rangle}{\left(\varepsilon_n-\varepsilon_{n'}\right)^2},
\label{Berry}
\end{equation}
where $v_{x,y} = \partial H/\partial k_{x,y}$. Numerically diagonalizing the Hamiltonian Eq.~(\ref{H}) and computing 
the Chern number 
we find that 
\begin{equation}
\sigma_{xy} = \left\{\begin{array}{c} 4e^2/h\;\mathrm{sgn}(M),\,\,\,\,\,U<M \\
0,\,\,\,\,\,\,\,\,\,\,\,\,\,\,\,\,\,\,\,\,\,\,\,\,\,\,U>M
%\mathcal{C} = \left\{\begin{array}{c} 4\;\mathrm{sgn}(M),\,\,\,\,\,U<M \\
%0,\,\,\,\,\,\,\,\,\,\,\,\,\,\,\,\,\,\,\,\,\,\,\,\,\,\,U>M
\end{array}\right..
\label{BilayerHall}
\end{equation}
For $U < M$,
the Chern number is twice that of the single-layer graphene case,
corresponding to four edge modes. The bilayer graphene 
system behaves as an quantum anomalous Hall insulator when $U < M$, 
and exhibits vanishing Hall effect when $U > M$. The gated bilayer
graphene system therefore has a Hall current which is tunable by an external gate voltage. 

The potential difference $U$ breaks the bilayer's top-bottom spatial inversion symmetry.
This produces a valley Hall effect in
which valley-resolved electrons scatter to opposite sides of the
sample. This can be characterized by the valley Hall conductivity
which is defined as the difference between
the valley-resolved Hall conductivities $\sigma_{xy}^{\mathrm{v}} =
\sigma_{xy}^{\mathrm{K}}-\sigma_{xy}^{\mathrm{K'}}$.  We find that  
\begin{equation}
\sigma_{xy}^{\mathrm{v}} =\left\{\begin{array}{c} 4e^2/h\;\mathrm{sgn}(U),\,\,\,\,\,U>M \\
0,\,\,\,\,\,\,\,\,\,\,\,\,\,\,\,\,\,\,\,\,\,\,\,\,\,\,U<M
%\mathcal{C}_v =\left\{\begin{array}{c} 4\;\mathrm{sgn}(U),\,\,\,\,\,U>M \\
%0,\,\,\,\,\,\,\,\,\,\,\,\,\,\,\,\,\,\,\,\,\,\,\,\,\,\,U<M
\end{array}\right.. \label{BilayerVHall}
\end{equation}
Therefore, despite having a vanishing Chern number when $U > M$, the system exhibits a finite valley Hall conductivity
$4e^2/h$. The quantum anomalous Hall and quantum valley 
Hall effects thus occupy complementary regions in the $U-M$ phase
space, as summarized in the phase diagram Fig.~\ref{fig1}. The gated
bilayer graphene system therefore behaves, depending on whether $U$ or $M$ is larger, either as a quantum anomalous Hall
insulator or a quantum valley Hall insulator. 

\section{IV. Edge State Properties}

We have presented an analysis of bulk topological properties in
single-layer and bilayer graphene cases using the low-energy
Dirac Hamiltonian.  In this section, we study the
corresponding edge state properties on a finite single-layer and 
bilayer graphene sheet, and switch to a tight-binding representation
from which we obtain the edge bands numerically. 
%We note that 
%Refs.~\cite{Sandler}-\cite{Smith} have studied the band structures of
%single-layer and bilayer graphene nanoribbons with Rashba SO coupling,
%but without an exchange field. 
The finite-size single-layer and
bilayer graphene sheets in our calculations are terminated with zigzag
edges along one direction, and are infinite in the other direction.  
The Hamiltonian for single-layer graphene case can be expressed as
\begin{eqnarray}
H_{\mathrm{SLG}} &=& t \sum_{\langle{ij}\rangle \alpha }{ c^\dagger_{i
\alpha}c_{j\alpha}+ {i} t_{\mathrm{R}}\sum_{\langle{ij}\rangle \alpha \beta
}(\bm{s}_{\alpha
\beta}{\times}{\mathbf{d}}_{ij}){\cdot}\hat{\mathbf{z}}\,c^\dagger_{i \alpha } c_{j \beta }} \nonumber \\
&+& M
\sum_{i\alpha}{c^\dagger_{i\alpha}(s_{z})_{\alpha\alpha}c_{i\alpha}}, \label{SLGTB}
\end{eqnarray}
where the first term describes hopping between nearest-neighbors $i,j$ on the honeycomb lattice, the second
term is the Rashba SO term with coupling strength $t_{\mathrm{R}}$
($\mathbf{d}_{ij}$ is a unit vector pointing from site $j$ to site $i$), and the third term is the
exchange field $M$. $\alpha,\beta$ denote spin
indices. 

The bilayer graphene case is described by the Hamiltonian
\begin{eqnarray}
H_{\mathrm{BLG}} &=& H_{\mathrm{SLG}}^{\mathrm{T}}+H_{\mathrm{SLG}}^{\mathrm{B}}
+t_\bot \sum_{i\in\mathrm{T},j\in\mathrm{B},
\alpha} c^\dagger_{i\alpha}c_{j\alpha} \nonumber \\
&&+ U \sum_{i\in\mathrm{T},\alpha}{
c^\dagger_{i\alpha}c_{i\alpha}}-U \sum_{j\in\mathrm{B},\alpha}{
c^\dagger_{j\alpha}c_{j\alpha}},
\label{BLGTB}
\end{eqnarray}
%
% Allan:  Something more explicit has to be said about interlayer tunneling.  
% Check to see that my attempt is okay.  
% James: Ok
where $H_{\mathrm{SLG}}^{\mathrm{T, B}}$ are the top (T) and bottom
(B) layer Hamiltonians of  Eq.~(\ref{SLGTB}), 
vertical hopping $t_\bot$ between the layers is represented by the third 
term and occurs only between Bernal stacked neighbors, and the last two terms describe the potential difference $U$ applied
across the bilayer. The parameters in the tight-binding Hamiltonians
Eqs.~(\ref{SLGTB})-(\ref{BLGTB}) above are related to those in the
low-energy Hamiltonians Eq.~(\ref{eq1}) and Eq.~(\ref{H}) as $v =
3ta/2$ ($a = 1.42\,\AA$ is the graphene lattice constant), $\alpha = 2 t_{\mathrm{R}}/3$, 
and $M,U,t_\bot$ are the same in both equations.

\begin{figure}[!tb]
\includegraphics[width = 8.5cm,angle = 0]
%[width=8.5cm,totalheight=6.5cm,angle=0]
{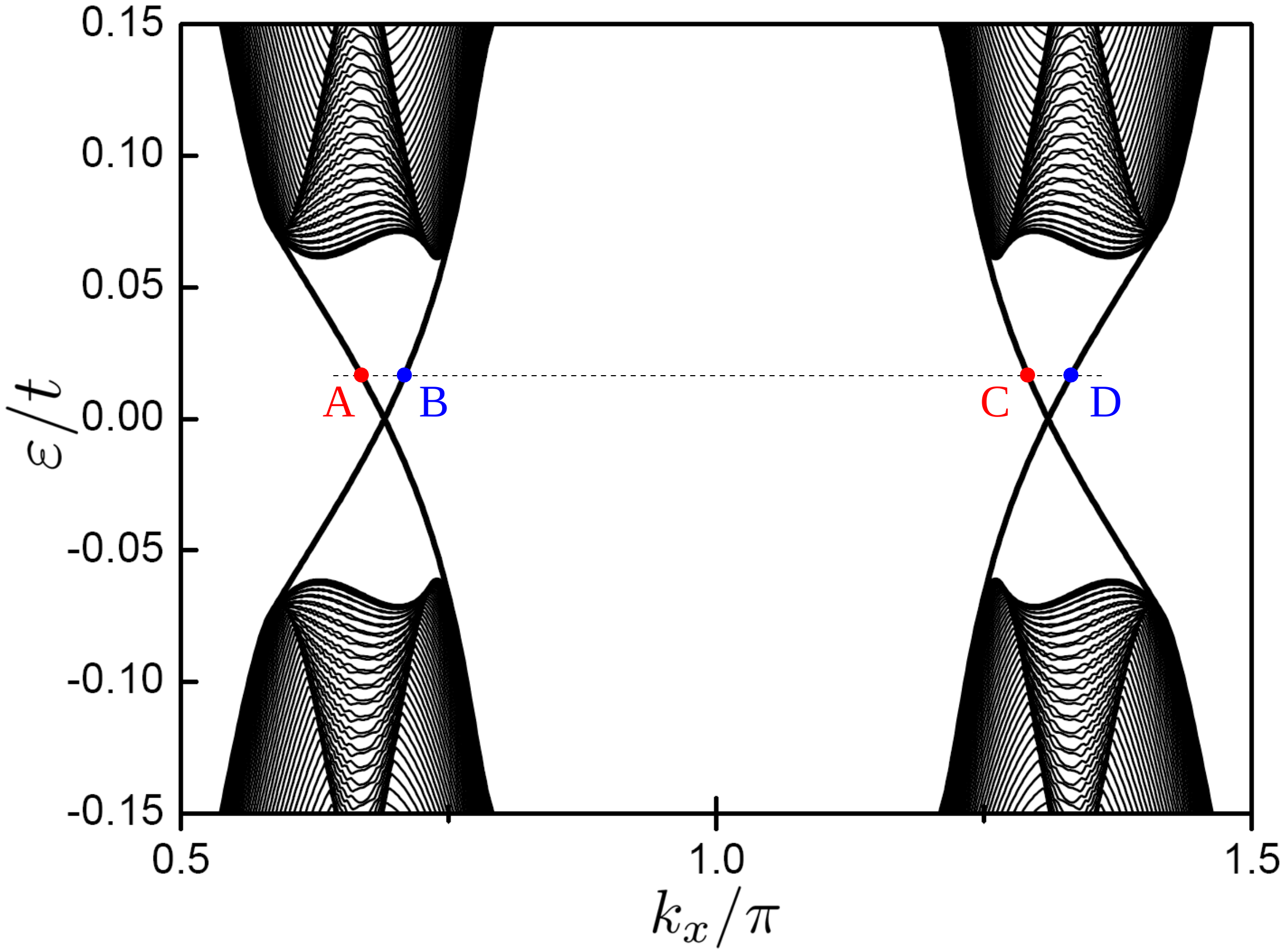}
\caption{(Color online) Edge state band structure of a single-layer graphene 
 ribbon with $M=0.1885$ and $t_{\mathrm{R}}=0.0471$ in units of the near-neighbor hopping
amplitude $t$ in the tight-binding Hamiltonian
  Eq.~(\ref{SLGTB}). 
% $M=0.1885$ (i.e., $0.02\times 3 \pi$) and $t_{so}=0.0471$(i.e., $0.005\times 3 \pi$).
These values correspond to $M/\varepsilon_{\mathrm{c}} = 0.02$ and
$\alpha/\varepsilon_{\mathrm{c}} = 7.5\times10^{-3}$
%0.005$ 
in the low-energy
Hamiltonian Eq.~(\ref{eq1}).} 
\label{Band-Single1}
\end{figure}

\subsection{Single-layer graphene case}

\begin{figure}[!tb]
\includegraphics[width = 8.5cm,angle = 0]
%[width=8.5cm,totalheight=7.cm,angle=0]
{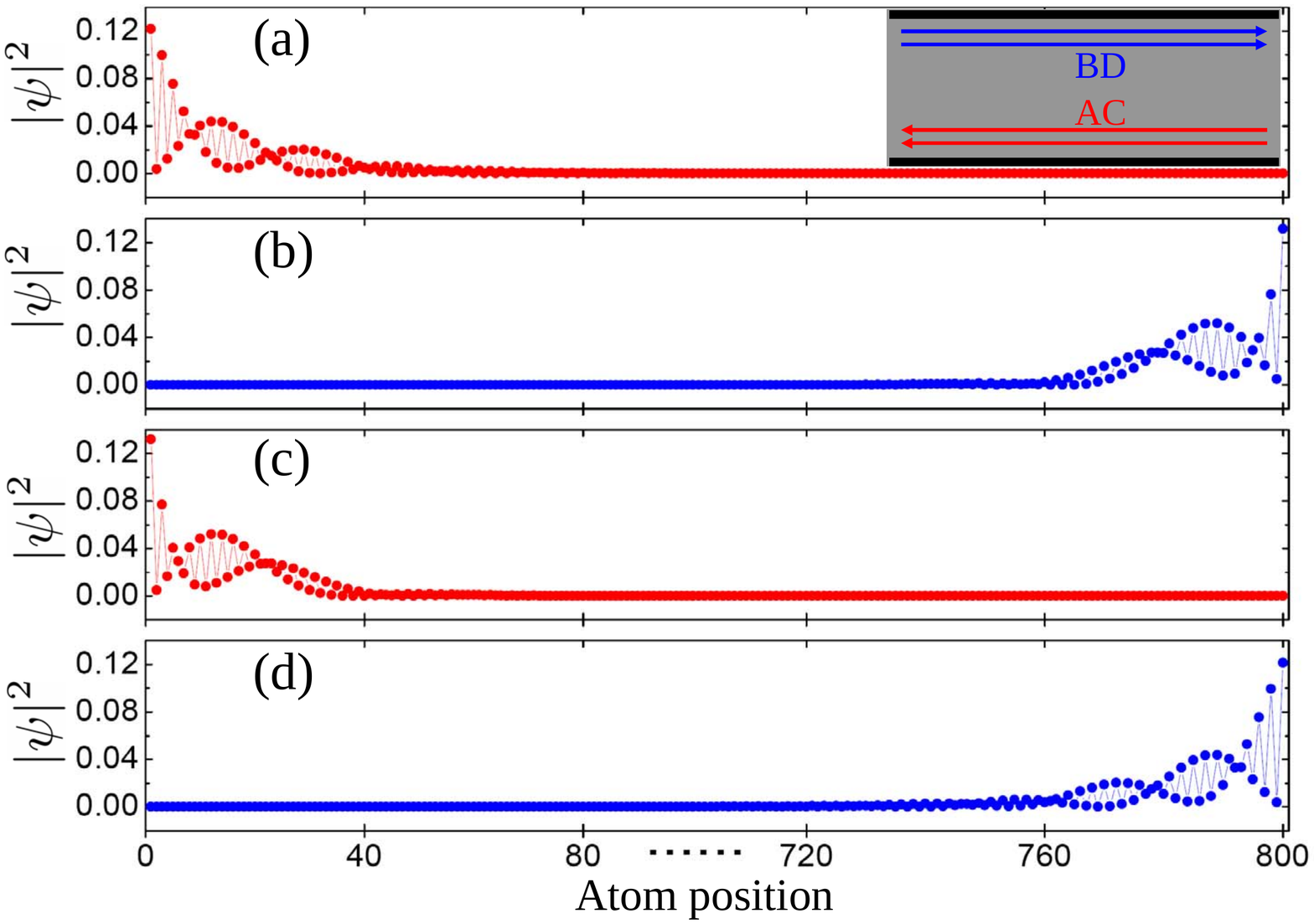}
\caption{(Color online) (a)-(d): Probability density  across the 
single-layer graphene sheet for the edge state wave functions 
$|\psi|^2$ of the edge states labeled A, B, C, D in 
Fig.~\ref{Band-Single1}. 
%(800 atoms per unit cell). 
The inset is a schematic which indicates he propagation
direction of the corresponding edge modes.}
\label{WaveFunction1}
\end{figure}

Fig.\ref{Band-Single1} shows the ribbon band structures calculated 
from Eq.~(\ref{SLGTB}). Inside the bulk gap, we find counter propogating gapless edge 
channels at each valley that are localized on opposite edges of the graphene sheet. 
In the Fig.\ref{WaveFunction1}, we plot the probability density
profile of edge state wave functions $|\psi|^2$ as a function of the
atom positions along the width of the graphene sheet for the four edge states labeled by A, B, C, D in Fig.\ref{Band-Single1}. It can be seen that
states A and C are localized along the left edge whereas B and D are
localized along the right edge. The edge states labeled by A and C have the same velocity and propagate along the same direction 
along one edge [inset of Fig.\ref{WaveFunction1}(a)], whereas B and D have opposite velocity and propagate
along the other edge. The two chiral edge modes each carry a unit
conductance $e^2/h$ yielding a quantized Hall conductivity
$\sigma_{xy} = 2e^2/h$. In the appendix we also present an envelope
function analysis of the edge state properties. The edge
state band structure obtained with this approach is found to be in
excellent agreement with the tight-binding results. 

\subsection{Gated bilayer graphene case}

\begin{figure*}[!tb]
\includegraphics[width = 18cm,angle = 0]
%[width=8.5cm,totalheight=6.5cm,angle=0]
{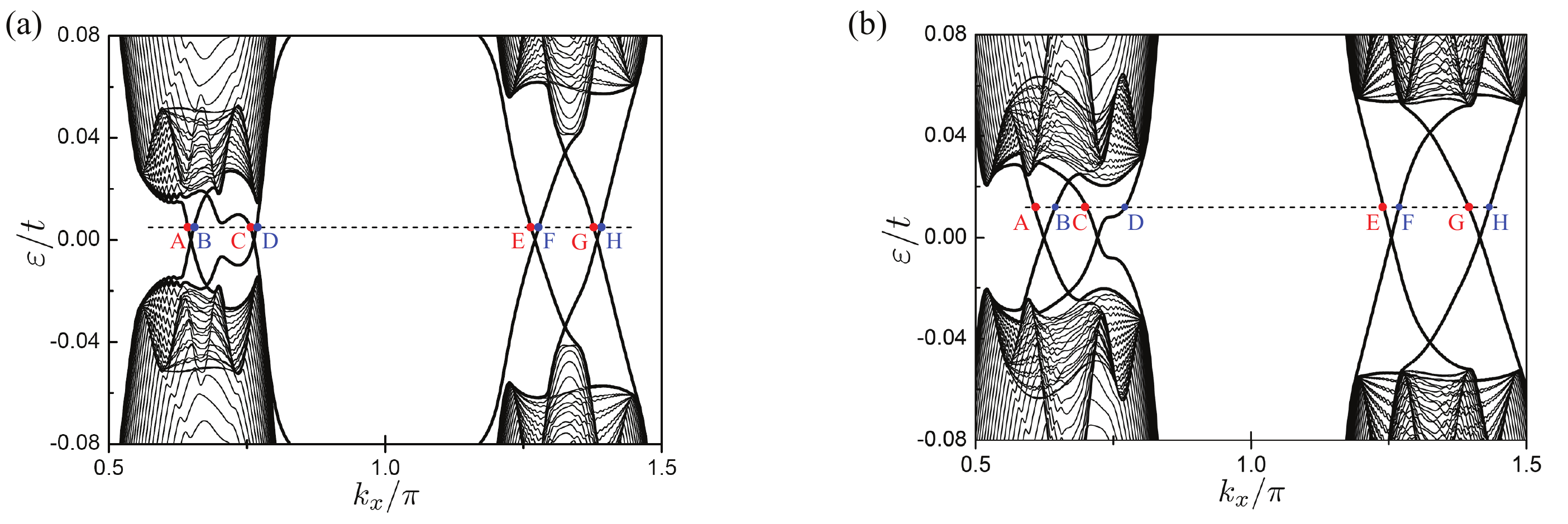}
%{band-bilayerMLB.pdf}
\caption{(Color online) Edge state band structure of bilayer graphene at fixed
Rashba SO coupling strength (in units of the hopping
amplitude $t$) $t_{\mathrm{R}} =0.0471$ for (a) $M > U$: $M=0.1885$ and
$U=0.0942$; (b) $U > M$: $M=0.0942$ and $U=0.2826$. These values correspond in the continuum 
Hamiltonian Eq.~(\ref{H}) $\alpha/\varepsilon_{\mathrm{c}} = 7.5\times10^{-3}$;
%0.005$; 
$M/\varepsilon_{\mathrm{c}} = 0.02$ and $U/\varepsilon_{\mathrm{c}} =
0.01$ for (a); $M/\varepsilon_{\mathrm{c}} = 0.01$ and $U/\varepsilon_{\mathrm{c}} =
 0.03$ for (b).} 
\label{band-bilayerMLB}
\end{figure*}
\begin{figure*}[!tb]
\includegraphics[width = 18cm,angle = 0]%[width=8.5cm,totalheight=6.5cm,angle=0]
{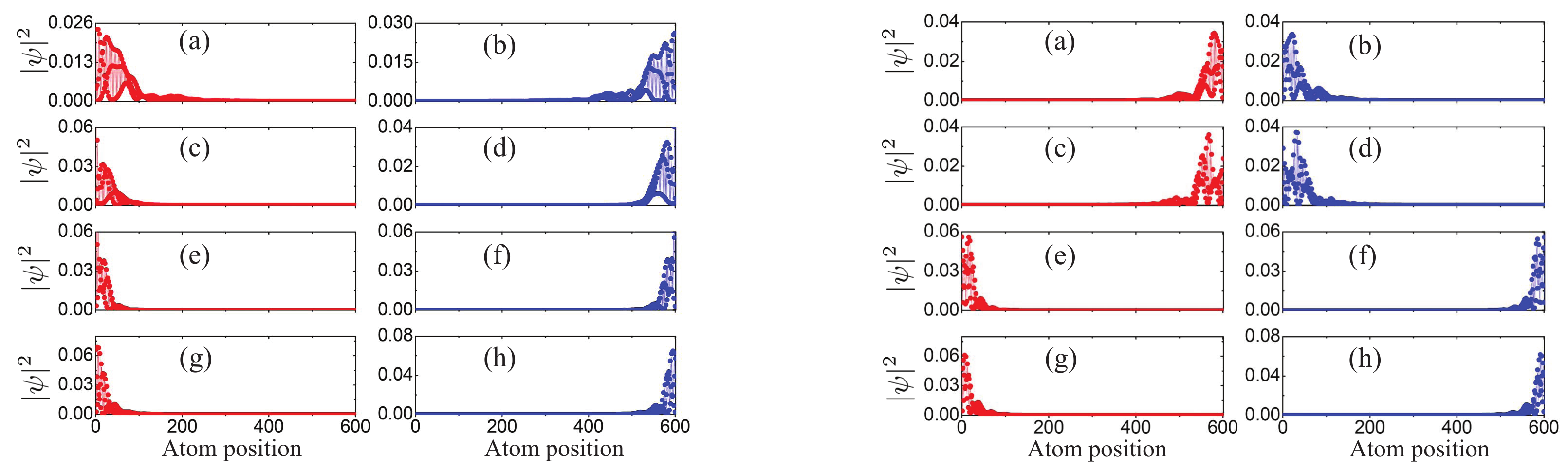}
%{Wave_Bilayer_MLB.pdf}
\caption{(Color online) (a)-(h): Probability density of the edge state wave function 
$|\psi|^2$ for the edge states A, B, C, D, E, F, G, H labeled in  
Fig.~\ref{band-bilayerMLB} for the case $M > U$ (left panel) and 
for $U > M$ (right panel) as a function of atom position. 
%on the top layer of the bilayer graphene sheet (400 atoms per unit
%cell for the top layer).
} \label{Wave_Bilayer_MLB}
\end{figure*}
\begin{figure}[!tb]
\includegraphics[width = 8cm,angle = 0]
%[width=8.5cm,totalheight=5cm,angle=0]
{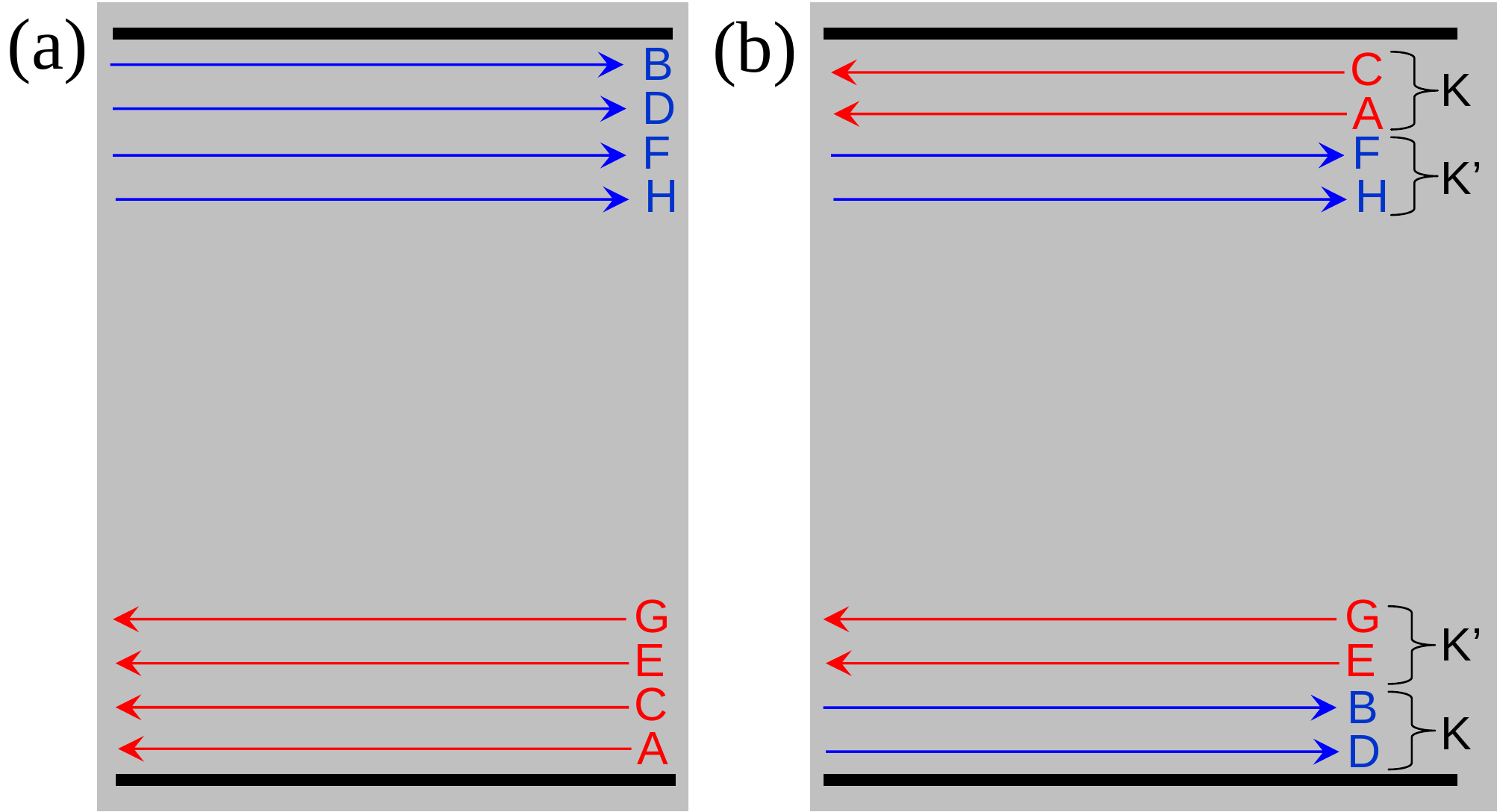}
\caption{(Color online) Schematic showing the direction of edge mode propagation
  (indicated by arrows) in the (a) quantum anomalous Hall phase; (b)
  quantum valley Hall phase.}
%Schematic plot of the edge states' propagation: (a) For the
%quantum Hall transport with exchange field being larger than the
%potential bias; (b) For the quantum valley Hall effect with exchange
%field being smaller than the potential bias.}
\label{Schematic_comparison}
\end{figure}

To study the edge state properties corresponding to the quantum
anomalous Hall phase and the quantum valley Hall phase, 
we show the edge state band structure at a fixed Rashba SO coupling
for the two cases $M > U$ and $U > M$ in Fig.~\ref{band-bilayerMLB}. 
 In contrast to the band structure in the single layer case 
Fig.~\ref{Band-Single1}, we find that the bilayer graphene band structure becomes asymmetric at K and K'. 
Within the bulk gap, there exists two edge bands associated with each
valley. In Fig.~\ref{Wave_Bilayer_MLB} we show the probability
density of the edge state wave function $|\psi|^2$ for the edge states labeled from A to H in
Fig.~\ref{band-bilayerMLB} for both cases. The left panel shows the
case $M > U$, and we find that the edge states labeled by A, C, E, G
are localized on one edge whereas B, D, F, H are localized on the
other edge. This corresponds to the quantum anomalous Hall phase
[Fig.~\ref{Schematic_comparison}(a)], 
where there exists four parallel chiral edge modes yielding a
quantized Hall conductance $\sigma_{xy}=4e^2/h$. 

For the case $U > M$, the right panel of the wave function plot Fig.~\ref{Wave_Bilayer_MLB} reveals that the four edge modes A,
C, E, G which propagate along the same direction now become split between
the two edges: A and C travel along one edge whereas E and G travel
along the opposite edge. Similarly, for the edge modes that travel in
the opposite direction, B and D propagate along one edge whereas E and
G propagate along the other edge. This is illustrated in the schematic
of Fig.\ref{Schematic_comparison}(b). 
Since the total current along one edge now adds up to zero, the Hall
conductivity vanishes. Nevertheless, two sets of
counter-propagating edge modes that belong to different valleys K and
K' travel along one edge. This situation bears a remarkable resemblance to
the quantum spin Hall effect where one edge consists of two
counter-propagating spin polarized modes. Due to the broken top-bottom
layer spatial inversion symmetry, bilayer graphene exhibits a quantized valley Hall effect, with
$\sigma_{xy}^{\mathrm{v}} = 4e^2/h$. In the case of single-layer
graphene, such a quantized valley Hall effect arises when the A-B
sublattice symmetry is broken; however there is no obvious strategy for 
imposing such an external potential experimentally. Through
top and back gating, bilayer graphene allows for a more experimentally
accessible way to produce the quantum valley Hall effect. Our present results show
that the quantum valley Hall effect can coexist with time reversal
symmetry breaking, provided that the breaking of spatial inversion
symmetry wins over that of time reversal symmetry.   

\section{IV. Conclusion}

In summary, we have studied the quantum anomalous Hall 
effect in single-layer and bilayer graphene systems with strong Rashba 
spin-orbit interactions due to externally controlled inversion symmetry breaking, 
and strong exchange fields due to proximity coupling to a 
ferromagnet. For neutral single-layer
graphene, we find that the Hall conductivity is quantized as 
$\sigma_{xy} = 2e^2/h$. For 
bilayer graphene, in which an external gate voltage can introduce an 
inversion symmetry breaking gap, we find a quantized Hall conductivity at 
neutrality equal to 
$4e^2/h$ when the potential difference $U$ is smaller than the exchange
% Allan:  Changed sign of inequality below.  Check.
coupling $M$.  This anomalous Hall effect is similar to the 
quantized anomalous Hall effect \cite{sahetheory,saheexpt} which can occur
spontaneously in high-quality 
bilayers at low-temperatures, but is potentially more robust because it relies on
external exchange and spin-orbit fields rather than spontaneously broken symmetries.
When $U > M$, the system exhibits a quantized valley
Hall effect with valley Hall conductivity $4e^2/h$. 
% By virtue of the two-dimensional nature of the single-layer and bilayer graphene sheet, an in-plane magnetic field
% might also be used in place of an exchange field produced by a
% ferromagnet. 
%
%

Two obstacles stand in the way of realizing the quantum anomalous Hall
effects discussed in this paper experimentally. It will be necessary 
first of all to introduce a sizeable Rashba spin-orbit coupling. One possibility is surface deposition of 
heavy-nucleus magnetic atoms that induce large
spin-orbit coupling.  The exchange field that is also
required could be introduced through a proximity effect. 
From our ab initio studies \cite{TBpaper}, an exchange field splitting of $56\,\mathrm{meV}$ and Rashba
spin-orbit coupling $2.8\,\mathrm{meV}$ can be obtained by depositing 
Fe atoms on the graphene surface. Another possible solution is to deposit graphene on a ferromagnetic
insulating substrate. The presence of a substrate breaks spatial inversion
symmetry, and therefore also produces Rashba spin-orbit coupling. 
Since the exchange field is induced through a proximity effect, 
layered antiferromagnetic insulators, which
are more abundant in nature can also be used and offer the advantage of an enlarged
pool of candidate substrate materials.   

%Secondly, one would also need to introduce an
%Zeeman exchange field to graphene. This could be accomplished 
%by using magnetic transition metal atoms that have heavy nuclei, so
%that a large spin-orbit coupling and 
%It will be necessary first of 
%all to introduce a sizeable Rashba spin-orbit coupling, possibly by 
%surface deposition of heavy-nucleus atoms that carry large
%spin-orbit coupling. Secondly, one would also need to introduce an
%Zeeman exchange field to graphene. This could be accomplished 
%by using magnetic transition metal atoms that have heavy nuclei, so
%that a large spin-orbit coupling and 
%
% {\em Can we suggest that it is possible to 
%use the same atom to create SO coupling and exchange coupling? 
%I think that there were some ideas in this direction mentioned in the proposal.} 
%{\em Add an extra paragraph discussing possible uses if this effect was realizable in practice.
%In principle Hall transport is not dissipative.   Is that useful?  You might want to read some of the 
%speculations that people have made concerning the spin Hall effect for inspiration.
%Also have people talked about applications of the quantum Hall effect at low temperatures.
%Perhaps edge transport is good?}  

\section{Acknowledgement}

This work was supported by the Welch grant  F1473 and by DOE
grant (DE-FG03-02ER45958, Division of Materials Sciences and 
Engineering). Y.Y. was supported by NSFC (No. 10974231) 
and the MOST Project of China (2007CB925000, 2011CBA00100).

\section{Appendix: Envelope function analysis of edge modes}

In this appendix, we present results for the edge state band
structures using an envelope function approach based on the continuum
Dirac model. We shall present only the single-layer graphene case
below, as the bilayer case does not offer as much analytic
tractibility since the Hamiltonian Eq.~(\ref{H}) is not analytically
diagonalizable. We first calculate the envelope-function band
structure and then compare the results with the tight-binding band structure. 

\begin{figure}[h]
  \includegraphics[width=8.5cm,angle=0]{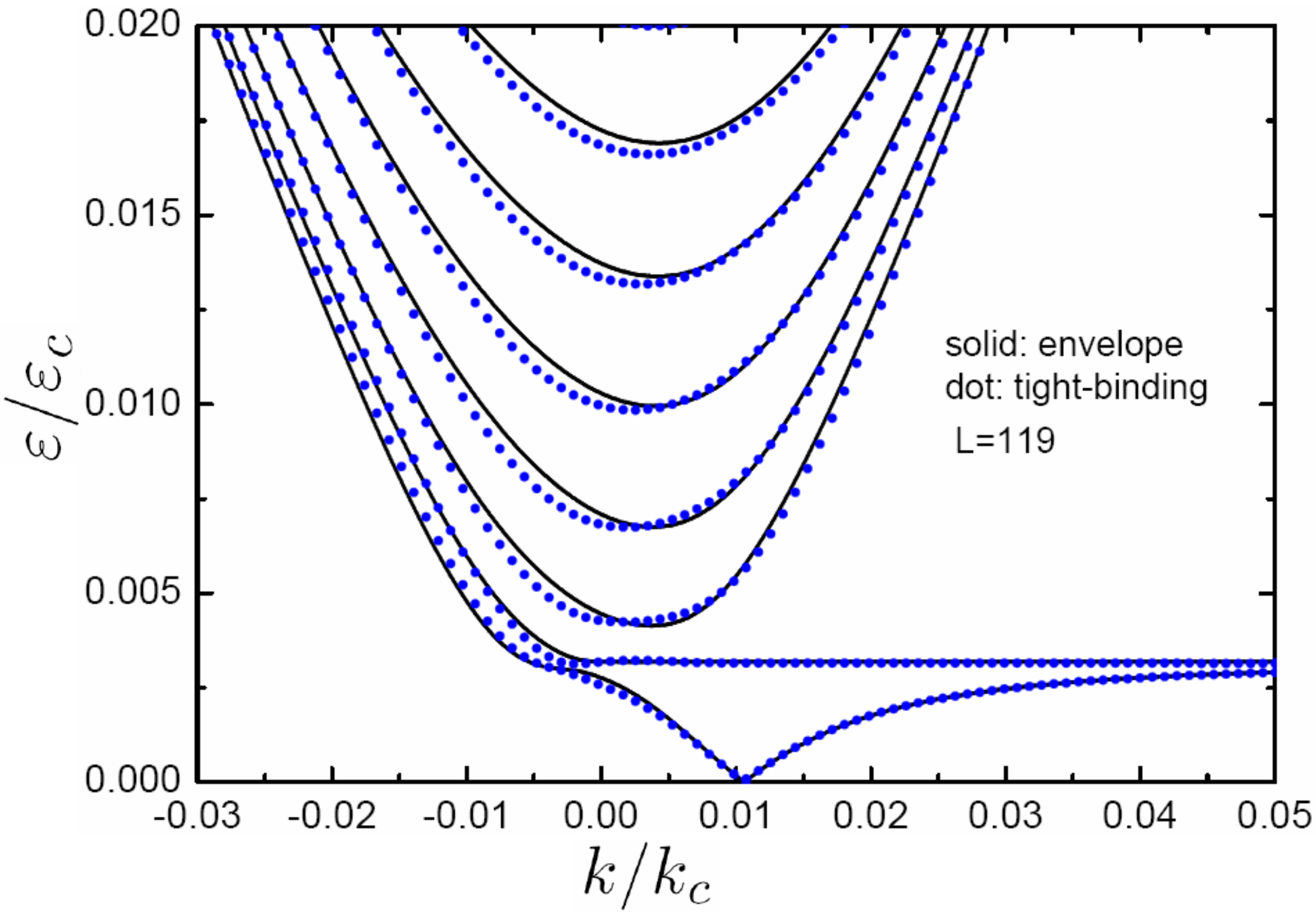}
%{Envelope_TB_Bands_Labels.pdf}
\caption{(Color online) Band structure from the envelope function approach
  and tight-binding model. The Rashba spin-orbit strength and exchange
  field in the tight-binding model are  (in units of $t$) 
  $t_{\mathrm{R}}=0.02$ and $M = 0.1$ in Eq.~(\ref{SLGTB}),
  corresponding to $\alpha/\varepsilon_{\mathrm{c}} = 2.12\times
  10^{-3}$ and $M/\varepsilon_{\mathrm{c}} = 0.01$ in the low-energy Dirac Hamiltonian.
% The Rashba spin-orbit strength $\alpha$ 
%   and exchange field $M$ in the envelope function approach, corresponding to the tight-binding parameter
%   (in units of $t$) 
%   $t_{\mathrm{R}}=0.02$ and $M = 0.1$ in Eq.~(\ref{SLGTB}), are
%   $\alpha/\varepsilon_{\mathrm{c}} = 2.122\times 10^{-3}$ and $M/\varepsilon_{\mathrm{c}} = 0.011$. 
  The width
  of the graphene sheet is $L = 119$ in units of the nearest-neighbor
  lattice constant. 
} \label{EnvelopeTB}
\end{figure}

Consider below a graphene sheet of infinite extent in the
$x$ direction but finite in the y direction spanning from $y = 0$ to
$y = L$. From the Hamiltonian Eq.~(\ref{eq1}), one can write down the
eigenvalue problem satisfied by the 
wavefunction $\tilde{\bm{\Psi}} = \left[\begin{array}{cccc}
\tilde{\psi}_{\mathrm{A}\uparrow}, &
\tilde{\psi}_{\mathrm{B}\uparrow}, & 
\tilde{\psi}_{\mathrm{A}\downarrow}, &
\tilde{\psi}_{\mathrm{B}\downarrow}\end{array}\right]^{\mathrm{T}}$
%satisfies the following eigenvalue problem 
%
\begin{eqnarray}
\left[\begin{array}{cccc}
-\varepsilon+M & v\left(k_x-\partial_y\right) & 0 & 0 \\
v\left(k_x+\partial_y\right) & -\varepsilon+M & -i 2
\alpha & 0 \\
0 & i 2 \alpha & -\varepsilon-M &
v\left(k_x-\partial_y\right) \\
0 & 0 & v\left(k_x+\partial_y\right) & -\varepsilon-M
\end{array}\right]\tilde{\bm{\Psi}} 
%\left[\begin{array}{c}
%\tilde{\psi}_{\mathrm{A}\uparrow} \\
%\tilde{\psi}_{\mathrm{B}\uparrow} \\ 
%\tilde{\psi}_{\mathrm{A}\downarrow} \\
%\tilde{\psi}_{\mathrm{B}\downarrow}\end{array}\right] \nonumber \\
= 0. \nonumber \\
\label{eq8}
\end{eqnarray}
For zigzag-edged graphene, the following boundary conditions apply
\begin{eqnarray}
\tilde{\psi}_{\mathrm{A}\uparrow}\left(y = L\right) &=&
\tilde{\psi}_{\mathrm{A}\downarrow}\left(y = L\right) = 0, \label{eq9} 
\\
\tilde{\psi}_{\mathrm{B}\uparrow}\left(y = 0\right) &=&
\tilde{\psi}_{\mathrm{B}\downarrow}\left(y = 0\right) = 0. \label{eq10} 
\end{eqnarray}
The solution of the problem Eq.~(\ref{eq8}) admits the ansatz
$\tilde{\bm{\Psi}} = e^{\lambda 
  y}\bm{\Psi}$. Substituting into Eq.~(\ref{eq8}), we obtain the
energy dispersion in terms of $\lambda$ from the resulting 
determinantal equation 
%$\tilde{\psi}_{\mathrm{A,B}\uparrow\downarrow} = e^{\lambda
%  y}\psi_{\mathrm{A,B}\uparrow\downarrow}$. 
%It follows that
%
%\begin{eqnarray}
%\left[\begin{array}{cccc}
%-\varepsilon+M & v\left(k_x-\lambda\right) & 0 & 0 \\
%v\left(k_x+\lambda\right) & -\varepsilon+M & -i 2
%\alpha & 0 \\
%0 & i 2 \alpha & -\varepsilon-M &
%v\left(k_x-\lambda\right) \\
%0 & 0 & v\left(k_x+\lambda\right) & -\varepsilon-M
%\end{array}\right]\left[\begin{array}{c}
%\psi_{\mathrm{A}\uparrow} \\
%\psi_{\mathrm{B}\uparrow} \\ 
%\psi_{\mathrm{A}\downarrow} \\
%\psi_{\mathrm{B}\downarrow}\end{array}\right] \nonumber = 0. \\
%\label{eq11}
%\end{eqnarray}
%
%We therefore obtain the energy in terms of $\lambda$ from the
%determinantal equation of Eq.~(\ref{eq11}):
%
%\begin{widetext}
\begin{eqnarray}
\varepsilon &=& 
\mu\left\{M^2+v^2\left(k_x^2-\lambda^2\right)+2\alpha^2\right. \nonumber \\
&&\left.+2s\sqrt{\alpha^4+v^2\left(k_x^2-\lambda^2\right)\left(M^2+\alpha^2\right)}\right\}^{1/2},
\label{eq12}
\end{eqnarray}
%\end{widetext}
%
%where $\mu = \pm$, $s = \pm$. 
which in turn yields four characteristic lengths
$\pm\lambda_{1,2}$ in terms of the energy
$\varepsilon$: 
\begin{equation}
\lambda_{1,2} =
\frac{1}{v}\sqrt{v^2k_x^2-\left[\varepsilon^2+M^2\pm2\sqrt{\varepsilon^2
      M^2+\alpha^2\left(\varepsilon^2-M^2\right)}\right]}.
\label{eq13}
\end{equation}
%
%corresponding to the decay lengths from the edge along the $y$ direction. 
Note that $\lambda_{1,2}$ in general can be complex, corresponding to a
mixture of the edge and bulk states. The eigenvectors can be obtained
as  
\begin{equation}
\bm{\Psi}\left(\lambda\right) 
%= \left[\begin{array}{c}
%\psi_{\mathrm{A}\uparrow} \\
%\psi_{\mathrm{B}\uparrow} \\ 
%\psi_{\mathrm{A}\downarrow} \\
%\psi_{\mathrm{B}\downarrow}\end{array}\right] 
=  
\left[\begin{array}{c}
-\left(\varepsilon+M\right)\left[v^2\left(\lambda^2-k_x^2\right)+\left(\varepsilon-M\right)^2\right]
\\
i2\alpha\left(\varepsilon^2-M^2\right) \\
i2\alpha v\left(\varepsilon+M\right)\left(k_x-\lambda\right) \\
-v\left(k_x+\lambda\right)\left[v^2\left(\lambda^2-k_x^2\right)+\left(\varepsilon-M\right)^2\right]
\end{array}\right],
\label{eq14}
\end{equation}
where we have left out an inessential normalization constant. The total
wavefunction can therefore be represented as a linear superposition of
the constituent basis wavefunctions 
\begin{eqnarray}
\bm{\tilde{\Psi}} &=& 
C_1 \bm{\Psi}\left(\lambda_1\right)e^{\lambda_1
  y}+D_1 \bm{\Psi}\left(-\lambda_1\right)e^{-\lambda_1 y} \nonumber \\
&&+C_2\bm{\Psi}\left(\lambda_2\right)e^{\lambda_2 y}+D_2
\bm{\Psi}\left(\lambda_2\right)e^{-\lambda_2 y}.
\label{eq15}
\end{eqnarray}
Using the boundary conditions Eqs.~(\ref{eq9})-(\ref{eq10}), we obtain
the following determinantal equation
\begin{widetext}
\begin{equation}
\begin{vmatrix}
f(\lambda_1) e^{\lambda_1
  L}  &
f(\lambda_1) e^{-\lambda_1
  L} &
f(\lambda_2) e^{\lambda_2
  L} &
f(\lambda_2) e^{-\lambda_2 L}
\\
1 & 1 & 1 & 1 \\
\left(k_x-\lambda_1\right)e^{\lambda_1 L} & \left(k_x+\lambda_1\right)e^{-\lambda_1 L} &
\left(k_x-\lambda_2\right)e^{\lambda_2 L} &
\left(k_x+\lambda_2\right)e^{-\lambda_2 L} \\
f(\lambda_1)\left(k_x+\lambda_1\right)
&
f(\lambda_1)\left(k_x-\lambda_1\right)
&
f(\lambda_2)\left(k_x+\lambda_2\right)
&
f(\lambda_2)\left(k_x-\lambda_2\right)
\end{vmatrix} = 0.
\label{eq16}
\end{equation}
\end{widetext}
where $f(\lambda) =
v^2\left(\lambda^2-k_x^2\right)+\left(\varepsilon-M\right)^2$. With
$\lambda_{1,2}$ given by Eq.~(\ref{eq13}), the band structure 
$\varepsilon$ as a function of $k_x$ can be obtained from solving
Eq.~(\ref{eq16}). We illustrate in Fig.~\ref{EnvelopeTB} the resulting band structure in the
vicinity of a Brillouin zone corner, from which it can be seen that both the bulk and
edge bands obtained from the envelope function approach show
excellent agreement with the tight-binding result. 
*On leave from University of Texas at Austin.


\begin{thebibliography}{18}
\bibitem{Remark} One is tempted to arrive at a $4\times4$ low-energy
  Hamiltonian from the $8\times8$ Hamiltonian Eq.~(\ref{H}) by treating the
  tunneling perturbatively as in Ref.~\cite{Mccann}. Such a Hamiltonian however would not
  predict the correct quantized anomalous Hall or valley Hall
  conductivity, because the higher-energy bands also contribute to the 
  Hall response.
\end{thebibliography}
\end{document}